\documentclass{article}

\usepackage{latexsym}
\usepackage{graphicx}
\usepackage{amsmath}

\begin{document}

\begin{centering}
 \large \textbf{Using weighting algorithms to refine source direction determinations in all-sky gravitational wave burst searches with two-detector networks II: The case of elliptical polarization} \\
\vspace{0.25cm}
\normalsize
\noindent T. McClain\footnote{Correspondence to: tjamcclain2@gmail.com} \\
Weinberg Theory Group, University of Texas, Austin, Texas, USA \\
\end{centering}

\renewcommand{\abstractname}{}
\begin{abstract}
I expand on the results of a recent work in which a novel weighting algorithm was shown to substantially increase the accuracy of an old, non-Bayesian computational approach for inferring the source direction of a gravitational wave from the output of a two-detector network. While that work was limited to the consideration of circularly polarized gravitational waves, the current analysis shows that the same approach is even more successful when applied to the generic case of elliptically polarized gravitational waves.
\end{abstract}

\vspace{0.5in}

\noindent \textbf{PACS.}
04.80.Nn \, Gravitational wave detectors and experiments -- 95.55.Ym \, Gravitational radiation detectors; mass spectrometers; and other instrumentation and techniques
 
\section{Introduction}
\label{intro}
In a recent paper \cite{mcclain2018using} I suggested using weighting algorithms to revitalize a very old method of determining the source location of gravitational waves in all-sky burst searches with two detector networks. The method begins by simply simulating a great many possible signals and looking for the one that produces the smallest deviation from the actual detector responses, but improves on the performance of that single-best-fit method by positing that, in the absence of any single excellent fit -- as we must expect in the presence of substantial noise or if we do not wish to expend the extraordinary computational resources required to produce accurate results with high-performance Bayesian analysis methods like LALInference -- it should be the sky angles that produce the largest number of better-than-average fits that represent the true source location. I showed in \cite{mcclain2018using} that there are invariably a number of weighting algorithms that allow this revised method to improve on the results a single-best-fit analysis in the cases I studied. However, due to computational restrictions, I only applied the new algorithms to circularly polarized gravitational waves or on un-realistically small (i.e., not all-sky) searches. This paper extends the analysis to the full parameter space of monochromatic all-sky burst searches.

None of the motivations or fundamental assumptions have changed since \cite{mcclain2018using}, and I have tried not to spent too much time re-iterating the basic goals of the approach or where it stands relative to other methods; I refer the interested reader to the original work instead. However, for the sake of completeness I note a few important works not cited in my original paper. \cite{becsy2017parameter} details a recent method that uses a minimal assumptions model similar to \cite{mcclain2018using}, but within a Bayesian analysis framework that extracts sky location as well as other parameters. \cite{klimenko2011localization} and more recently \cite{essick2015localization} detail the general criteria by which all-sky burst sources are localized.

I have re-iterated the technical details upon which the method rests, especially in section \ref{sec:methods}. The reader who is familiar with these details from \cite{mcclain2018using} can safely skip to the second paragraph of section \ref{sec:results} for the details and results of the new simulations.

\section{Methods}
\label{sec:methods}

To extend my previous analysis, I continue to build upon the basic approach used in \cite{mcclain2018using}. As before, I model the incoming waves with a sine-Gaussian waveform, as used by the LIGO collaboration in their all-sky burst search event detection algorithms \cite{abbott2017all}. The success of a particular parameter set \( \Theta \) is quantified by summing the absolute value of the difference between the algorithm's calculated responses \( R_{out} \) and the simulated detector responses \(R_{in}\) for each detector in the network \( N \) and each sampled time  \( t \in T \) in the lifetime of the signal: 
\begin{equation} 
Q(\Theta) := \sum_{n \in N \, , \, t \in T} \sqrt{ \left(R_{n,out}(t, \Theta) - R_{n,in}(t) \right)^2 }
\label{eq:ellipticquality}
\end{equation} 
Though the more common choice would be to normalize to the noise variance:
\begin{equation}
Q'(\Theta) := \sum_{n \in N \, , \, t \in T} \sqrt{ \left( \frac{(R_{n,out}(t, \Theta) - R_{n,in}(t))^2}{\eta^2_{n}(t)} \right) }
\end{equation}
where the term $\eta_n(t)$ represents the (estimated) noise in the $n^{th}$ detector at time $t$, this is once again unnecessary, as this paper deals only with monochromatic signals and the noise is identical across all modeled parameter sets\footnote{This would not be the case if non-monochromatic signals were being analyzed after Fourier decomposition.}. 

The detector responses \( R_n(t) \) are calculated in the standard way (see, for example, \cite{schutz2011}) and include both randomly generated noise and other, non-random but un-modeled contributions to the waveform. Following the conventions used by Schutz in \cite{schutz2011}, we compute the response functions
\begin{equation}
R_{n} (t) = h_+(t + \tau_n) F^+_n (\theta, \phi) + h_{\times} (t + \tau_n) F^{\times}_n (\theta, \phi) +\eta_n(t) 
\label{eq:ellipticdetectorresponse}
\end{equation}
where \( h_+ \) and \( h_{\times} \) represent the two independent polarization amplitudes of the incoming gravitational wave (including both modeled and un-modeled contributions), \( \tau_n = \frac{1}{c} (\vec r_0 - \vec r_n) \cdot \hat e_{gw} \) represents the time delay between the \( n^{th} \) detector and an arbitrarily chosen ``\( 0^{th} \)'' reference detector, \( F^+_n \) and \( F^{\times}_n \) represent the beam pattern response functions of the \( n^{th} \) detector (that is, the response of the \( n^{th} \) detector to a unit-amplitude, linearly polarized signal \( h_+ = 1 \) or \( h_{\times} = 1 \) ), and \( \eta_n \) represents the noise in the \( n^{th} \) detector. Because actual gravitational wave detectors seem to have instrumental noise that does not necessarily match the Gaussian noise model (see, for example, \cite{abbott2016characterization}), this analysis assumes non-Gaussian noise. This noise is random-number generated, and is characterized throughout the paper by its maximum allowed value within a given set of simulations, \( \eta_{max} \), which is in turn set by the signal-to-noise ratio chosen for each simulation set: \( \frac{\sqrt{h^2_{+,max} + h_{\times, max}^2}}{\eta_{max}} = \text{SNR} \). The noise values in each detector are generated independently, and each is uniformly distributed within the range \( [ - \eta_{max}, \eta_{max} ] \) \footnote{This particular noise model is chosen primarily to signal the understanding that real detector noise is non-Gaussian. A more careful treatment of the instrument noise is necessary for a definitive analysis, as is a treatment that assumes Gaussian noise if this method is to be directly compared to other, similar approaches. Both will be included in future work.}.

For the purposes of this algorithm, the beam pattern response functions are fully general (see, for example, \cite{schutz2011} or \cite{anderson2001}, though I follow different angle conventions than the latter source). I modeled only monochromatic, sine-Gaussian signals of the form

\begin{equation}
h_{+ \, , \, \times} (t, q, \omega, a_n) = \exp (- q^2 t^2) (a_1 \cos \omega t + a_2 \sin \omega t)
\label{ellipticsinegaussian}
\end{equation}
though with the potential for the ``real'' (simulated) signal \( h_+ \, , \, h_{\times} \) to be modified by an arbitrary, un-modeled function, which the simulations include up to fifth order in \( t \) :
\begin{multline}
h_{+ \, , \, \times} (t, q, \omega, a_n, u_n) = \\ \exp (- q^2 t^2) (a_1 \cos \omega t + a_2 \sin \omega t) (1+u_1 t + u_2 t^2 + u_3 t^3 + u_4 t^4 + u_5 t^5)\
\label{eq:ellipticums}
\end{multline}
These un-modeled additions to the simulated signal do not conform to the sine-Gaussian model and therefore cannot be readily fit by the algorithm. They are included to simulate the fact that real sources that do not perfectly fit the simple sine-Gaussian model, even in the absence of instrumental noise. As with the noise, these un-modeled signal amplitudes are random-number generated and are characterized throughout the paper by their maximum allowed values within a given set of simulations. The fitting algorithms can easily be made to handle non-monochromatic signals after Fourier decomposition at the expense of greater computational cost; as before, I have avoided these extra computational costs in this analysis. 

The key difference between my weighting algorithm and the single-best-fit algorithm is that it deals with many different fits with non-minimal \( Q \) values in the final determination of the ``best fit'' sky angles. Specifically, it allows searches other than the single-best-fit found by maximizing the weighting function 

\begin{equation}
W = 
\begin{cases}
1 & \mbox{if} \ Q = Q_{min} \\
0 & \mbox{otherwise} \\
\end{cases} 
\label{eq:ellipticminQ}
\end{equation}
The weighting algorithm of eq. \eqref{eq:ellipticminQ} looks for the parameter set with the minimum value of \( Q \). My algorithm instead produces a weighted best fit after summing over all parameters with which we are not concerned (that is, everything except the sky angles) by minimizing an arbitrary weighting function \( W = f (Q/Q_{min}) \)
\begin{equation}
\min \left( \sum_{\Theta \setminus (\theta , \phi)} f \left( \frac{Q(\Theta)}{Q_{min}} \right) \right)
\end{equation}
The output of the algorithm consists of the \( F^+ \), \( F^{\times} \), and \( \tau \) parameters that result from the \( (\theta, \phi ) \) values of the parameter set that minimizes this weighted sum.

As in \cite{mcclain2018using}, I do not attempt to reconstruct the sky angles \( \theta \) and \(\phi \) directly. Rather, it is \( F^+_n(\theta, \phi) \), \( F^{\times}_n(\theta, \phi) \), and \( \tau_n(\theta, \phi) \) that determine whether a particular set of sky angles accurately determines the response functions measured in the detector network. Any sky angles that produce the same values of \( F^+_n(\theta, \phi) \) and \( F^{\times}_n(\theta, \phi) \) and the same time delays \( \tau_n(\theta, \phi) \) will produce the same detector responses. The actual values of these sky angles can be recovered (non-uniquely) by finding the intersection of the multi-valued inverses of the beam pattern and time delay functions.

\begin{figure*}
\centering
\includegraphics[scale=1]{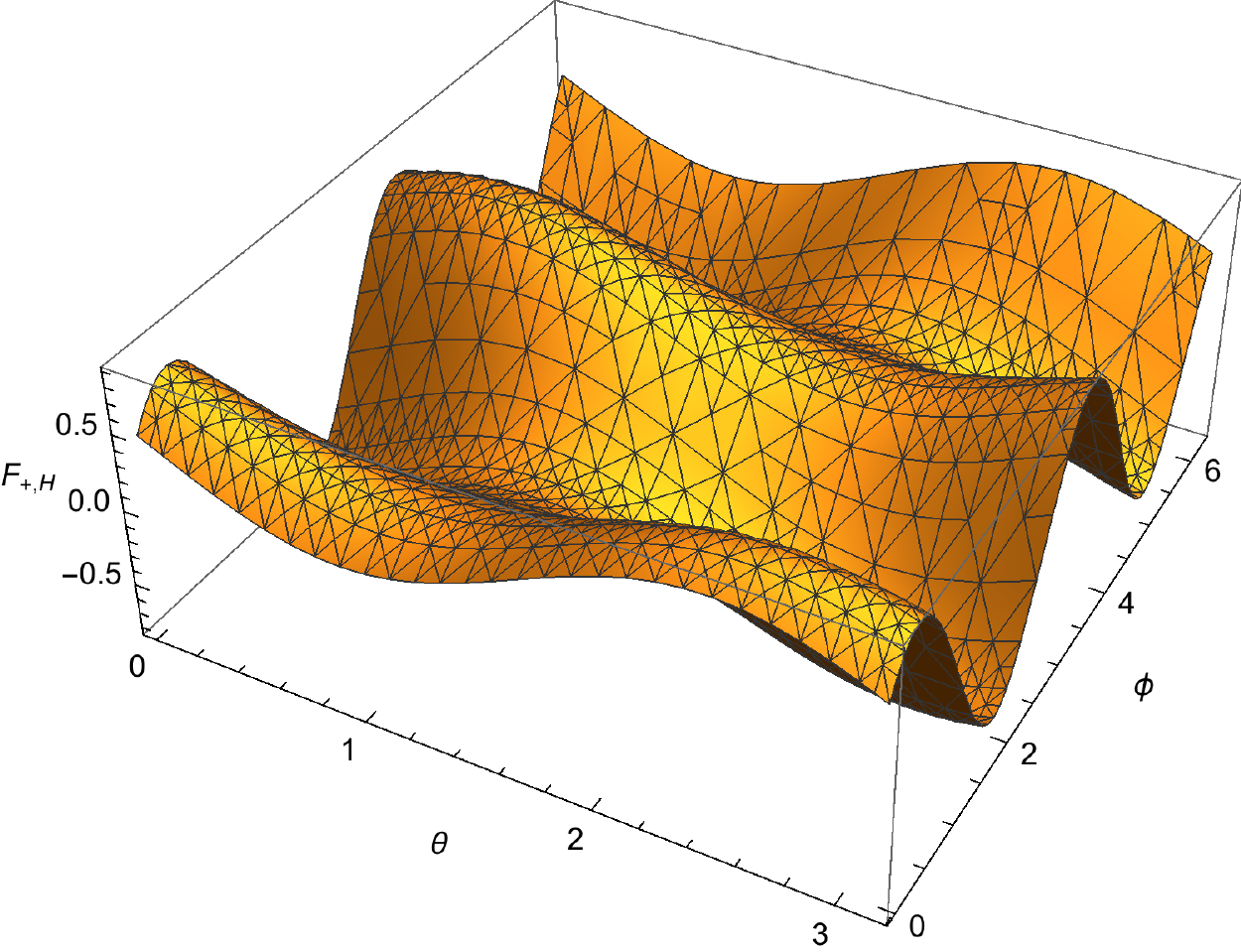}
\caption{This figure shows the beam pattern function $F_+$ for the LIGO-Hanford site at the reference value $\psi = 0$. The intersection between the graph and a horizontal plane intercepting the z-axis at the appropriate value gives the sky angles which yield a particular $F_+$; i.e., the multi-valued inverse. See, for example, \cite{schutz2011} for more information about how these beam pattern functions are calculated.}
\label{fig:fphplot}
\end{figure*}

\begin{figure*}
\centering
\includegraphics[scale=1]{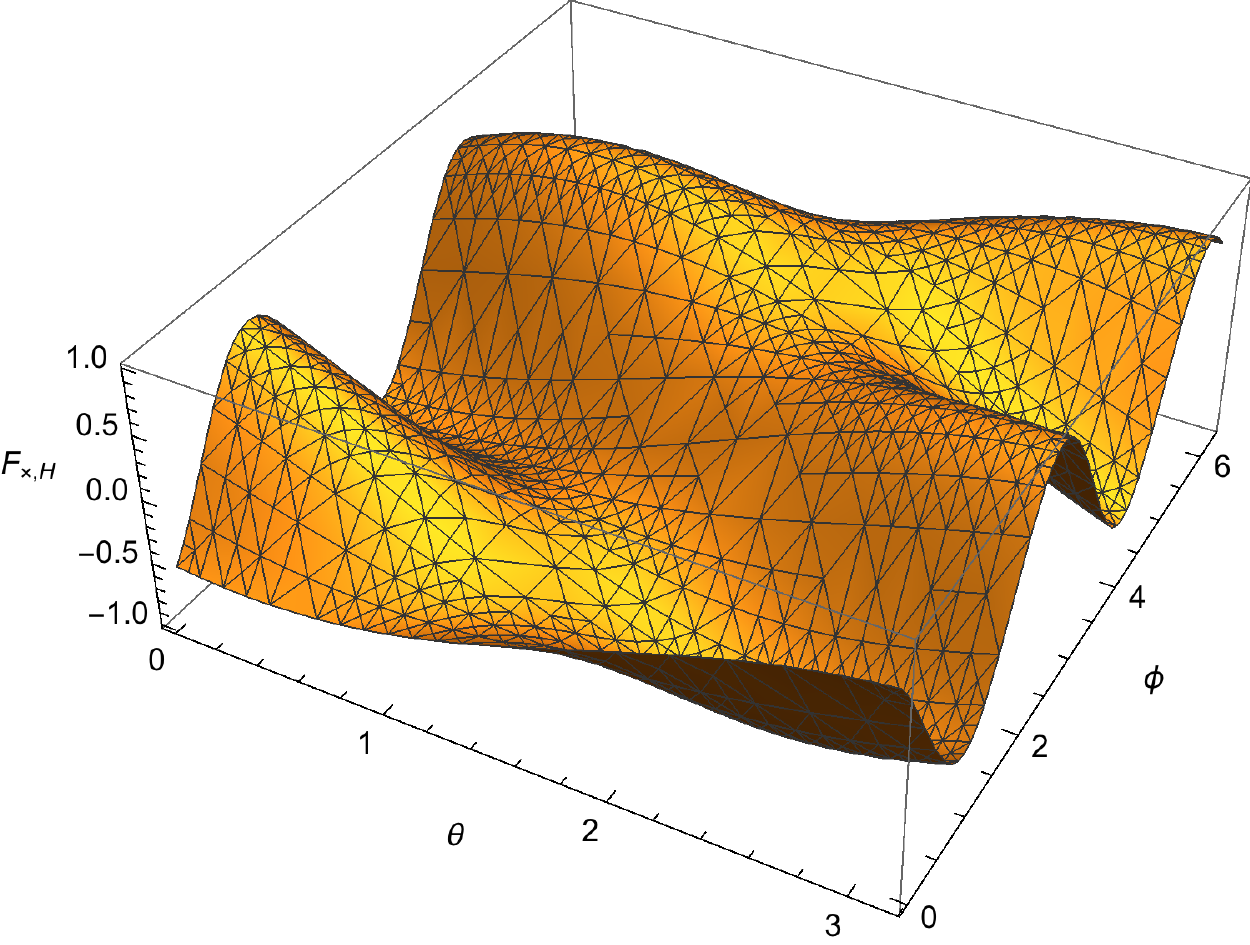}
\caption{This figure shows the beam pattern function $F_\times$ for the LIGO-Hanford site at the reference value $\psi = 0$. See the caption of Fig. \ref{fig:fphplot} for more information.}
\label{fig:fchplot}
\end{figure*}

\begin{figure*}
\centering
\includegraphics[scale=1]{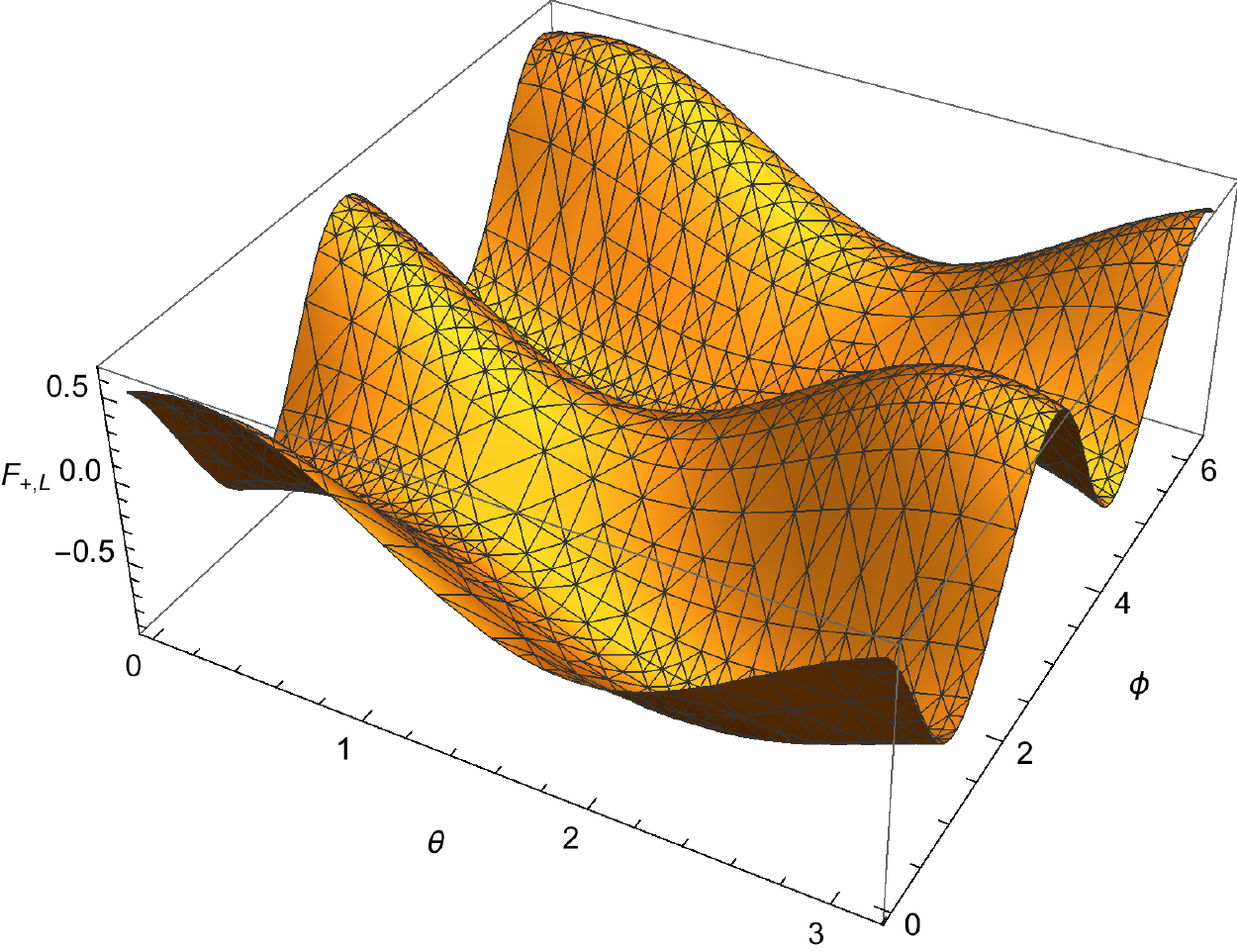}
\caption{This figure shows the beam pattern function $F_+$ for the LIGO-Livingston site at the reference value $\psi = 0$. See the caption of Fig. \ref{fig:fphplot} for more information.}
\label{fig:fplplot}
\end{figure*}

\begin{figure*}
\centering
\includegraphics[scale=1]{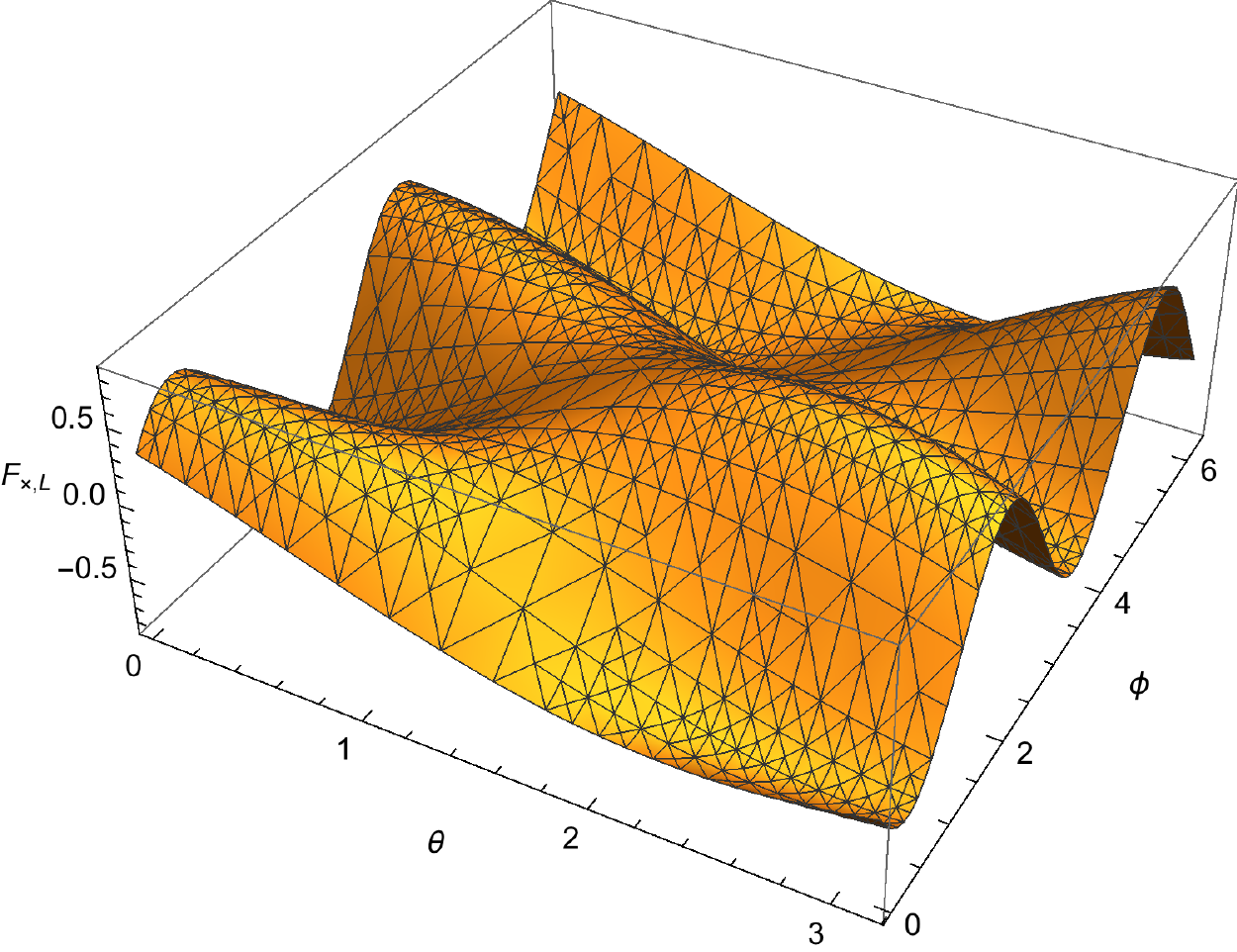}
\caption{This figure shows the beam pattern function $F_\times$ for the LIGO-Hanford site at the reference value $\psi = 0$. See the caption of Fig. \ref{fig:fphplot} for more information.}
\label{fig:fclplot}
\end{figure*}

\begin{figure*}
\centering
\includegraphics[scale=1]{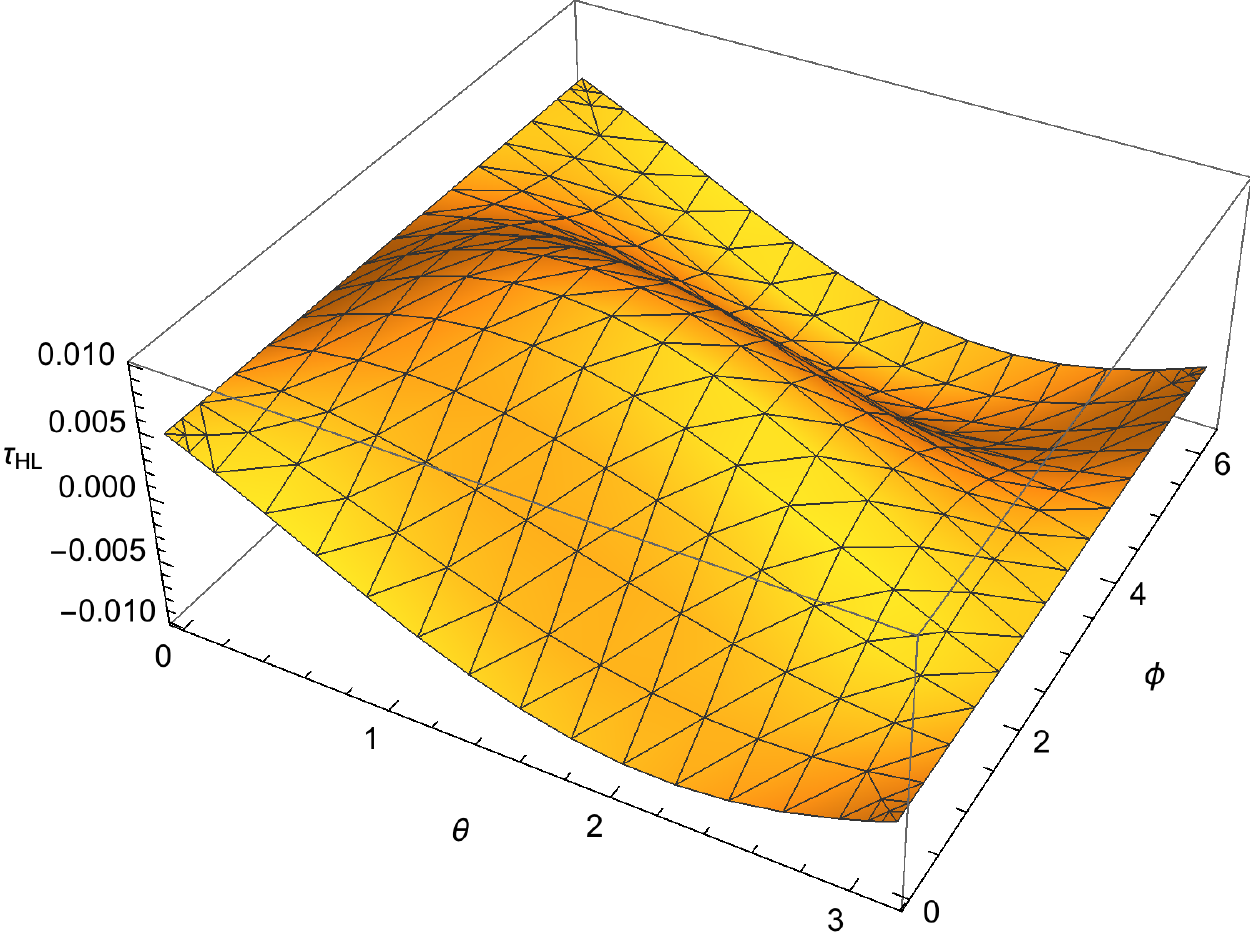}
\caption{This figure shows the time delay $\tau$ between the LIGO-Hanford and LIGO-Livingston sites. The intersection between the graph and a horizontal plane intercepting the z-axis at the appropriate value gives the sky angles which yield a particular $\tau$; i.e., the multi-valued inverse.  See the text immediately after eq. \eqref{eq:ellipticdetectorresponse} or \cite{schutz2011} for more information about how this time delay is calculated.}
\label{fig:tdhlplot}
\end{figure*}

To give a sense of just how multi-valued these beam pattern and time delay functions are, Figs. 
\ref{fig:fphplot} through \ref{fig:tdhlplot} show the beam pattern functions $F_+$ and $F_\times$ for each LIGO site for the sample value $\psi = 0$, as well as the time delay between the two sites. 

\section{Results}
\label{sec:results}

As in my analysis of circularly polarized gravitational waves, I characterize the performance of the various algorithms by the RMS difference between the model's predicted values for \( F^+ \) and \( F^{\times} \) at each site, the model's predicted \( \tau \) between the two sites, and the actual values of those five functions calculated from the randomly generated values \( ( \theta, \phi ) \) that produce the simulated signal. Values with \( in \) subscripts denote the values that serve as the algorithm's (simulated) input values (in place of real signal data), and \( out \) subscripts denote the values calculated from the parameter set identified by the algorithm as giving the best fit from among all the values randomly sampled from the parameter space:
\begin{multline}
\delta F_{rms} := \\ \frac{1}{4} \sqrt{(F^+_{1,out} - F^+_{1,in})^2 + (F^{\times}_{1,out} - F^{\times}_{1,in})^2 + (F^+_{2,out} - F^+_{2,in})^2 + (F^{\times}_{2,out} - F^{\times}_{2,in})^2}
\label{eq:ellipticfrms}
\end{multline}
\begin{equation}
\delta \tau_{rms} := \sqrt{ (\tau_{out} - \tau_{in})^2}
\label{eq:elliptictaurms}
\end{equation}
For specificity, I have chosen the two currently operating LIGO network detectors as the sites, with the Hanford detector set to be the reference site (\( \tau = 0 \)).

\begin{figure*}
\centering
\includegraphics[scale=1.25]{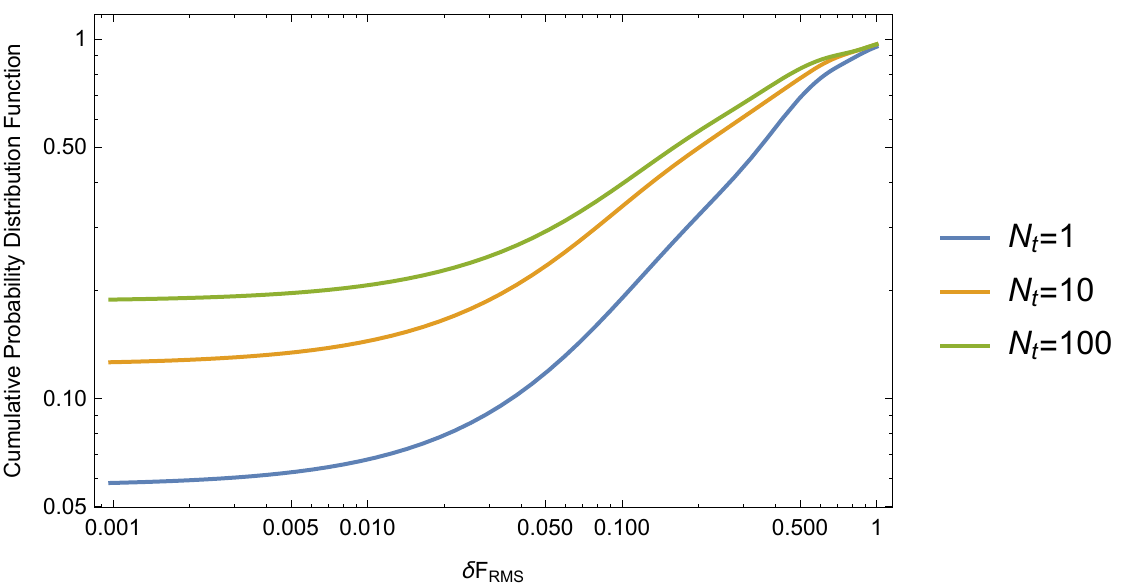}
\caption{This figure shows the substantial differences in fit RMS beam pattern response function error when the time portion of the parameter space is sampled more densely. The data indicates that the times are being appropriately sampled relative to source directions and GW amplitude combinations in the reference scenario. If times were being sampled much too densely then we would expect to see minimal change in the algorithm's ability to produce accurate predictions when the time sampling density was decreased. Conversely, if the times were being sampled much too sparsely then we would expect to see very large changes in the algorithm's ability to produce accurate predictions if they sampling density is increased. Instead, we see modest but significant changes in both scenarios. See section \ref{sec:results} for more details.}
\label{fig:elliptictsampleplot}
\end{figure*}

I continue to report the results of the somewhat specialized situation in which the frequency and q-factor of the incoming gravitational wave are known in advance, but the waves are no longer assumed to be circularly polarized. This is a reasonable approximation to situations in the signal is Fourier decomposed, and the analysis is performed on a handful of dominant frequencies within a handful of short time intervals. The parameter set used in generating the input signals for these simulations is: \(q \in \{ 2.15, 4.29, 8.58 \} \, \text{s}^{-1} \) (see \footnote{These seemingly random q-values produce signal lifetimes -- meaning times at which the signal drops to one-half its peak value -- of $1$s, $\frac{1}{2}$s, and $\frac{1}{4}$s, respectively}) \( f \in \{10, 100, 1000 \} \, \text{Hz} \) (see \footnote{The lower limit here pushes the state-of-the-art of what LIGO is capable of successfully detecting; we include it to show how the algorithm might behave in future GW detection scenarios}), \( \theta \in \left[ 0, \pi \right] \, \text{rad} \), \( \phi \in \left[ 0, 2 \pi \right] \, \text{rad} \), \( \text{SNR} \in \left[ 2, 100 \right] \), and the parameters \( u_1 \) through \( u_5 \) of eq. \eqref{eq:ellipticums} all capped at a single value \( u_{\text{max}} \in \left[ 1/100, 1/2 \right] \). The same ranges of values are used to generate the output signals, with the exception that the noise and un-modeled signal values have no analog in the algorithm's output signal calculations. I have chosen the particular parameter values \(q = 4.29 \, \text{s}^{-1} \), \( f =100 \, \text{Hz} \), \( \text{SNR} \geq 10 \), and the parameters \( u_1 \) through \( u_5 \) of eq. \eqref{eq:ellipticums} all capped at \( u_{\text{max}} = 1/10 \) as the baseline. In the figures that follow, most parameters are set to their baseline values, while a single parameter is changed to a new (specified) value. A graph that shows, for example, $f = 1000$ Hz has that given frequency, with all other parameter set to their baseline values. Each simulation is run with different random values for the input parameters, as well as freshly randomized noise and un-modeled signal, all within the appropriate bounds. Only a few times are checked over the lifetime of the signal (taken to be the time interval before it is suppressed by a factor of $2$ due to its exponential decay) due to computational restrictions (see fig. \ref{fig:elliptictsampleplot} below for more information about the impact of this approach). All graphs are the combined result of $1000$ simulations, each of which uses a single simulated signal (with noise as in eq. \eqref{eq:ellipticdetectorresponse} and un-modeled contributions as in eq. \eqref{eq:ellipticums} built in), but takes $N_t = 10$ random times, $N_{sd} = 100$ random source directions, and $N_{gwc} = 1000$ random gravitational wave amplitude combinations (the $a_1$ etc. from eq. \eqref{ellipticsinegaussian}) to generate model signals of the form of eq. \eqref{ellipticsinegaussian} to compare to the single simulated input signal.

Naturally, the performance of a revised algorithm with weighting function \( f (Q/Q_{min}) \) depends entirely upon the function \( f \). As in \cite{mcclain2018using}, I focus on weighting functions of the form
\begin{equation} 
W = \exp \left[ 1 - \left( \frac{Q}{Q_{min}} \right)^n \right]
\label{eq:ellipticweightedQ}
\end{equation}
However, in this larger parameter space, the specific values of $n$ that I chose to focus on lie in the new range \(n \in \left[ \frac{1}{64} , 4 \right] \). I began with the same, wider range reported in \cite{mcclain2018using}, but testing showed that the lower n-values above were generally more successful than the higher n-values I focus on in \cite{mcclain2018using}. All of these weighting functions still peak at \( Q = Q_{\min} \) with a value of \( 1 \), and reduce to exactly the result of eq. \eqref{eq:ellipticminQ} in the limit \( n \rightarrow \infty \). More intuitively, these functions look ``almost'' like the weighting function of eq. \eqref{eq:ellipticminQ}, but with some non-zero range over which ``good'' fits that are non-minimal can still contribute to the determination of the final angle values. For \( n = 2 \), the weighting function of eq. \eqref{eq:ellipticweightedQ} is a Gaussian centered on the minimum \( Q \) value. The baseline value $n = 2$ was used to generate all the graphs that follow because it represents a well-understood Gaussian weighting function, and gives smaller (i.e., better) average values than the single-best-fit approach for the $\delta F_{rms}$ and $\delta \tau_{rms}$ of eq. \eqref{eq:ellipticfrms} and \eqref{eq:elliptictaurms} that are used to characterize the accuracy of the algorithm's output in every scenario tested. However, in searching for the best possible weighting algorithms for specific parameter values, it is important to bear in mind that $n =2 $ was often not the exponent with the smallest average values in eq. \eqref{eq:ellipticfrms} and \eqref{eq:elliptictaurms}, representing a sensible baseline choice rather than a universal best result.

\begin{figure*}
\centering
\includegraphics[scale=1.25]{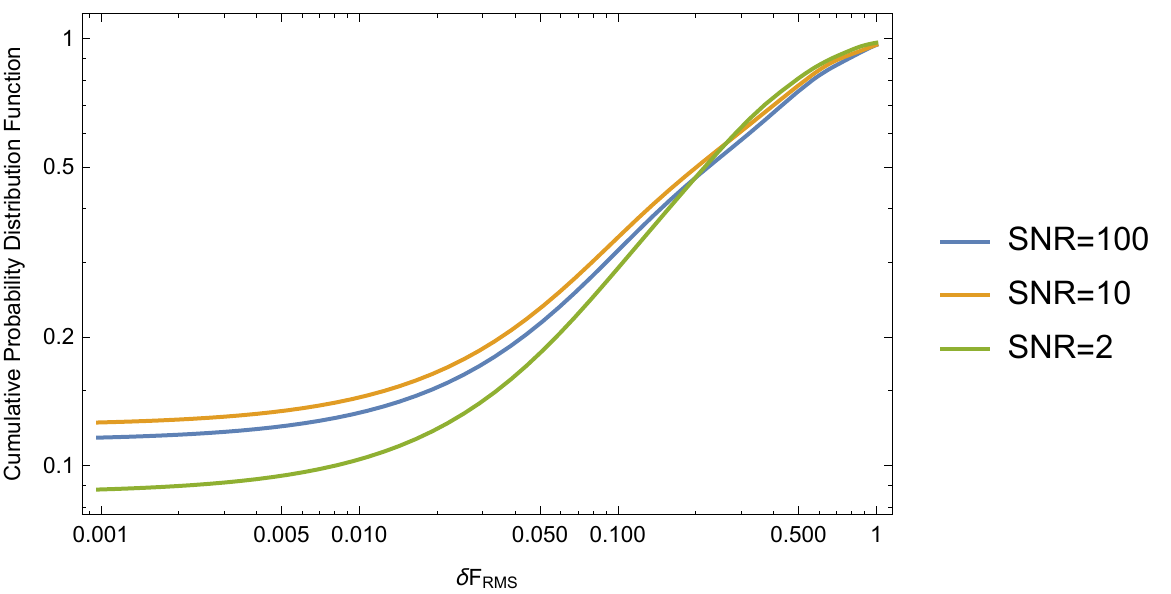}
\caption{This figure shows very minimal changes in the RMS beam pattern response function error across a very wide range of signal-to-noise ratios (see eq. \eqref{eq:ellipticdetectorresponse}) and fractional un-modeled signals (capped here at $\frac{1}{SNR}$; see eq. \eqref{eq:ellipticums}). Note, however, the clear decrease in the number of fits with $\delta F_{RMS} < ~0.1$ at SNR 2 compared to the others. These minimal changes should not be expected to persist when the parameter space is sampled more densely.}
\label{fig:ellipticsnrplot}
\end{figure*}

\begin{figure*}
\centering
\includegraphics[scale=1.25]{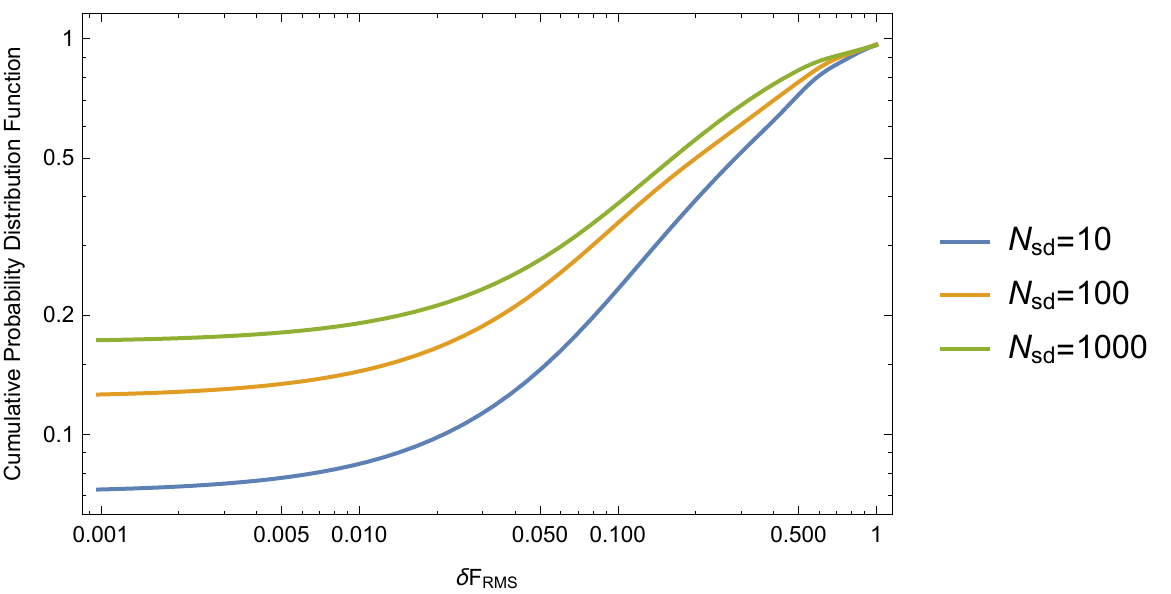}
\caption{This figure shows the substantial differences in RMS beam pattern response function error across the different numbers of source direction samples. As with the time sampling data, this indicates that the source directions are being appropriately sampled relative to times and GW amplitude combinations in the reference scenario. See section \ref{sec:results} for more details.}
\label{fig:ellipticanglesplot}
\end{figure*}

\begin{figure*}
\centering
\includegraphics[scale=1.25]{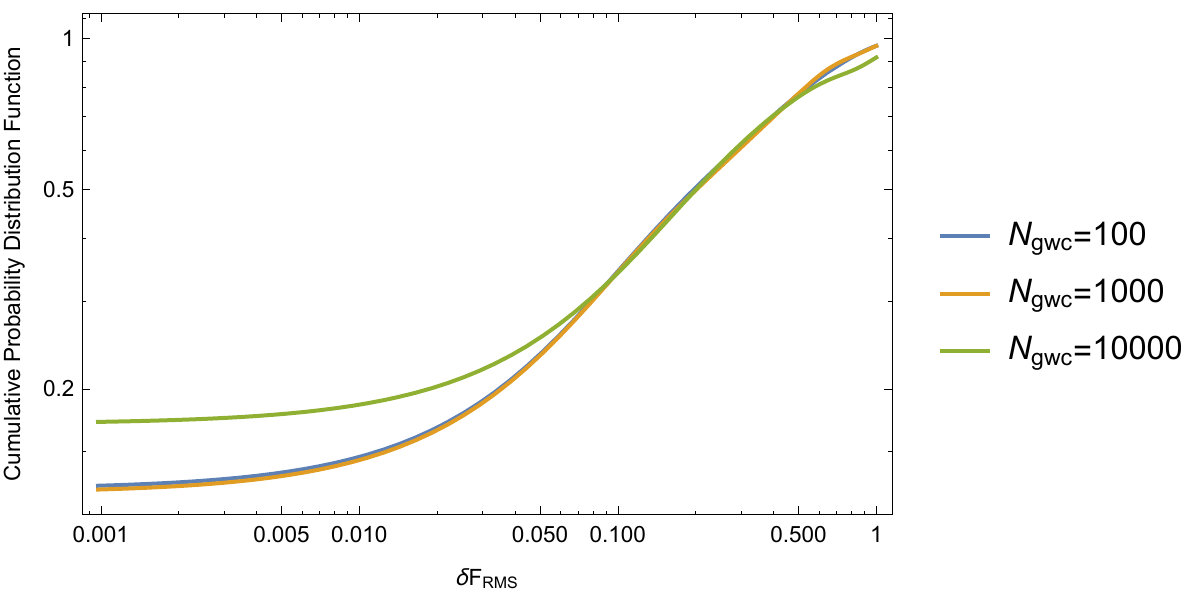}
\caption{This figure shows the very small difference in RMS beam pattern response function error across the different numbers of GW amplitude combinations. This data is ambiguous, indicating both that GW amplitude combinations are being over-sampled relative to times and source directions in the reference scenario, but also that the sampling density would need to be significantly higher to produce high-accuracy fit results. See section \ref{sec:results} for more details.}
\label{fig:ellipticgwcplot}
\end{figure*}

As expected, the weighting functions of eqs. \eqref{eq:ellipticweightedQ} and \eqref{eq:ellipticminQ} both perform substantially better -- even at very high noise levels and in the presence of substantial un-modeled signal -- than a ``random choice'' weighting function that simply weights every possible parameter value \( ( \theta, \phi ) \) the same. At every tested noise level there is one or more weighting function of the form of eq. \eqref{eq:ellipticweightedQ} that substantially improves upon the performance of the weighting function of eq. \eqref{eq:ellipticminQ}; see fig. \eqref{fig:ellipticsnrplot} for more details. Surprisingly, in this larger parameter space there are many scenarios in which \emph{every} tested value of $n$ improves upon the performance of the weighting function of eq. \eqref{eq:ellipticminQ}, suggesting that the method offers a substantial improvement of the key performance indicators $\delta F_{rms}$ and $\delta \tau_{rms}$ across a much wider range of weighting functions in this more realistic scenario. As noise decreases, the average value of both $\delta F_{rms}$ and $\delta \tau_{rms}$ decreases for the weighting functions of eq. \eqref{eq:ellipticminQ} and eq. \eqref{eq:ellipticweightedQ}. That the average values of the key performance indicators do not decrease more substantially at low noise levels is indicative of the fact that the parameter space is not being sampled densely enough to support many excellent fit results; see fig. \ref{fig:elliptictsampleplot}, \ref{fig:ellipticanglesplot}, and \ref{fig:ellipticgwcplot} to see the results of denser sampling.

In these graphs, the cumulative probability distribution (CPDF) function is plotted over the full possible range of $\delta F_{RMS}$ (see eq. \eqref{eq:ellipticfrms}). These line graphs show the fraction of all fits produced by the algorithm in these simulations that have a $\delta F_{RMS}$ below the value indicated on the horizontal axis. For example, the median value of $\delta F_{RMS}$ can be found by finding the horizontal intercept of the value $0.5$ on the CPDF; half of all fits had a $\delta F_{RMS}$ value lower than this. Table \ref{tab:ellipticmediandF} lists these median values for easy reference. More intuitively, these graphs give a sense of how accurately the algorithm is able to find the values of $F^+$ and $F^\times$ at each site.

\begin{table*}
\centering
\includegraphics[scale=1.25]{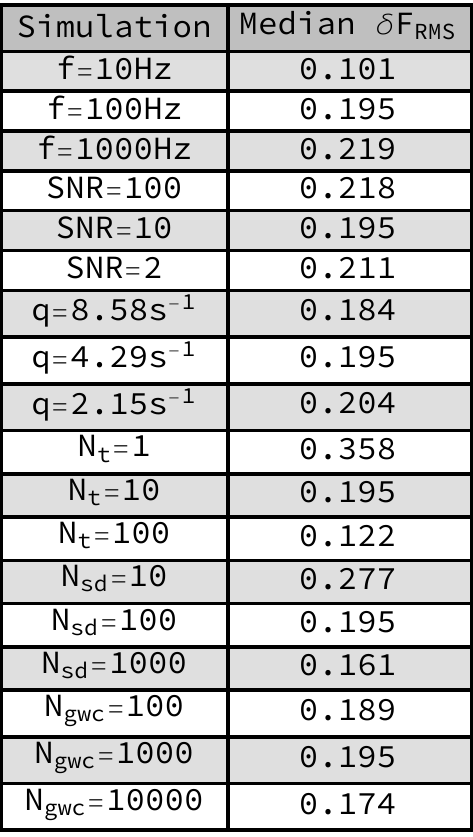}
\caption{This table shows the median value of $\delta F_{RMS}$ from eq. \eqref{eq:ellipticfrms} in each set of simulations. This is the same as the horizontal intercept of the value $0.5$ on each of the CPDF graphs shown below. Each simulation set is referenced by its single, non-baseline value. $N_t$ stands for the number of time samples, $N_{sd}$ for the number of source direction samples, and $N_{gwc}$ for the number of gravitational wave amplitude combinations. See section \ref{sec:results} for more details.}
\label{tab:ellipticmediandF}
\end{table*}

\begin{figure*}
\centering
\includegraphics[scale=1.25]{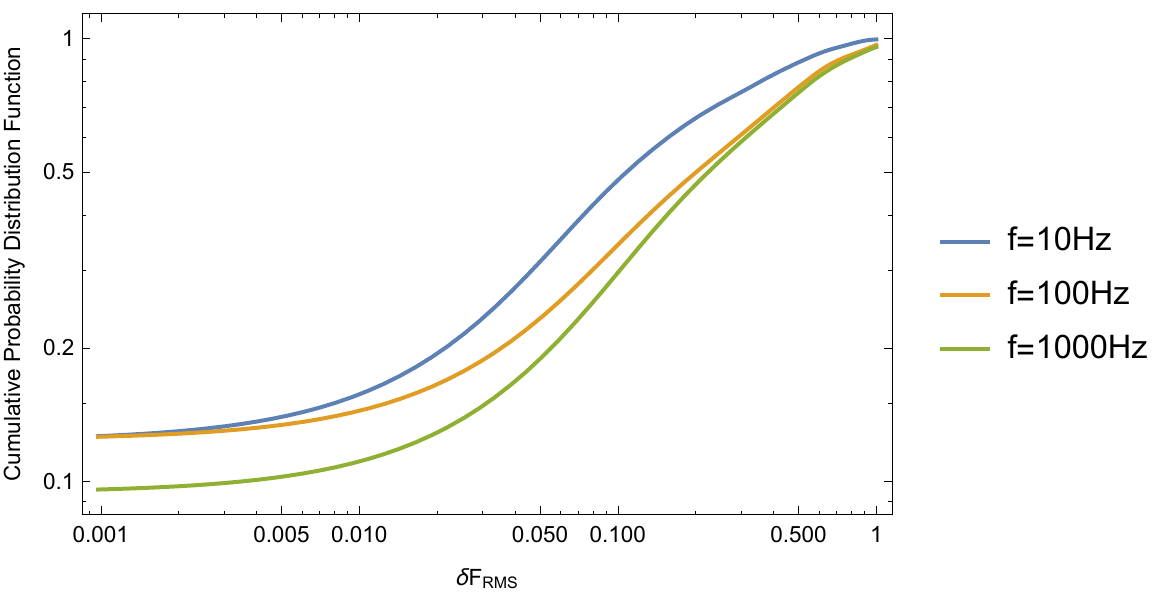}
\caption{This figure shows the relatively small changes in RMS beam pattern response function error across three orders of magnitude in frequency. The lowest frequency data -- near the limit of what LIGO can currently detect -- shows substantially better performance when compared to the two more realistic scenarios. As noted in the text, the CPDF here allows us to see what fraction of simulations produced a $\delta F_{RMS}$ (see eq. \eqref{eq:ellipticfrms}) less than the value on the horizontal axis. For example, the median value of $\delta F_{RMS}$ is the horizontal intercept of the value $0.5$ on the line graph. The scaling of the axes is logarithmic.}
\label{fig:ellipticfreqplot}
\end{figure*}

\begin{figure*}
\centering
\includegraphics[scale=1.25]{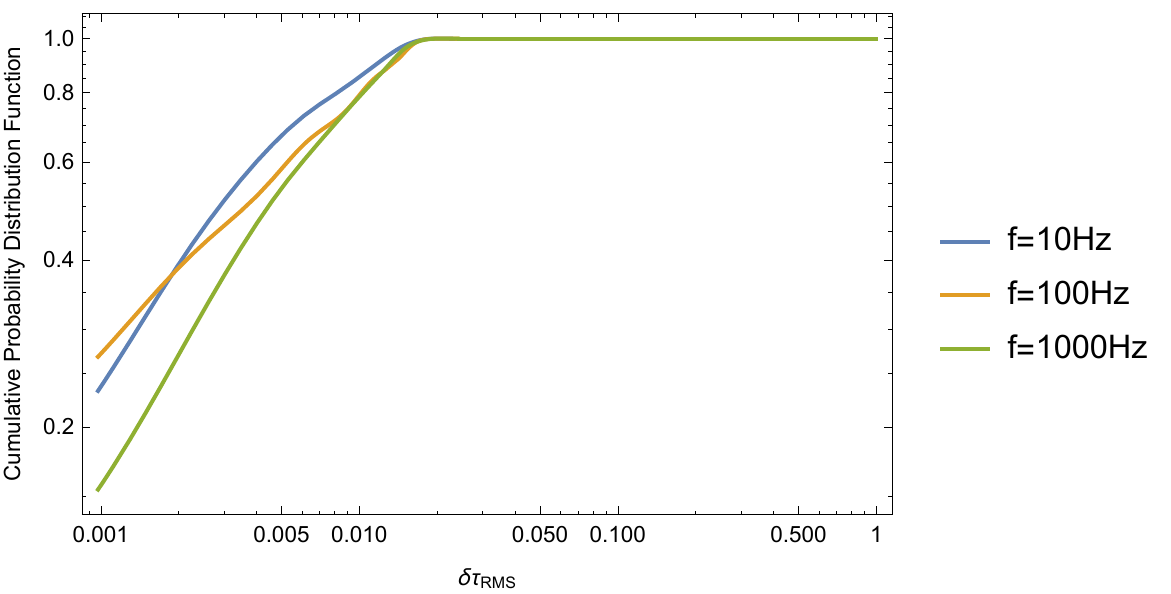}
\caption{This figure shows the relatively small changes in RMS time delay error across three orders of magnitude in frequency. The lowest frequency data -- near the limit of what LIGO can currently detect -- shows somewhat better performance when compared to the two more realistic scenarios. As noted in the text, the CPDF here allows us to see what fraction of simulations produced a $\delta \tau_{RMS}$ (see eq. \eqref{eq:elliptictaurms}) less than the value on the horizontal axis. For example, the median value of $\delta \tau_{RMS}$ is the horizontal intercept of the value $0.5$ on the line graph. The scaling of the axes is logarithmic.}
\label{fig:ellipticfreqtauplot}
\end{figure*}

\begin{figure*}
\centering
\includegraphics[scale=1.25]{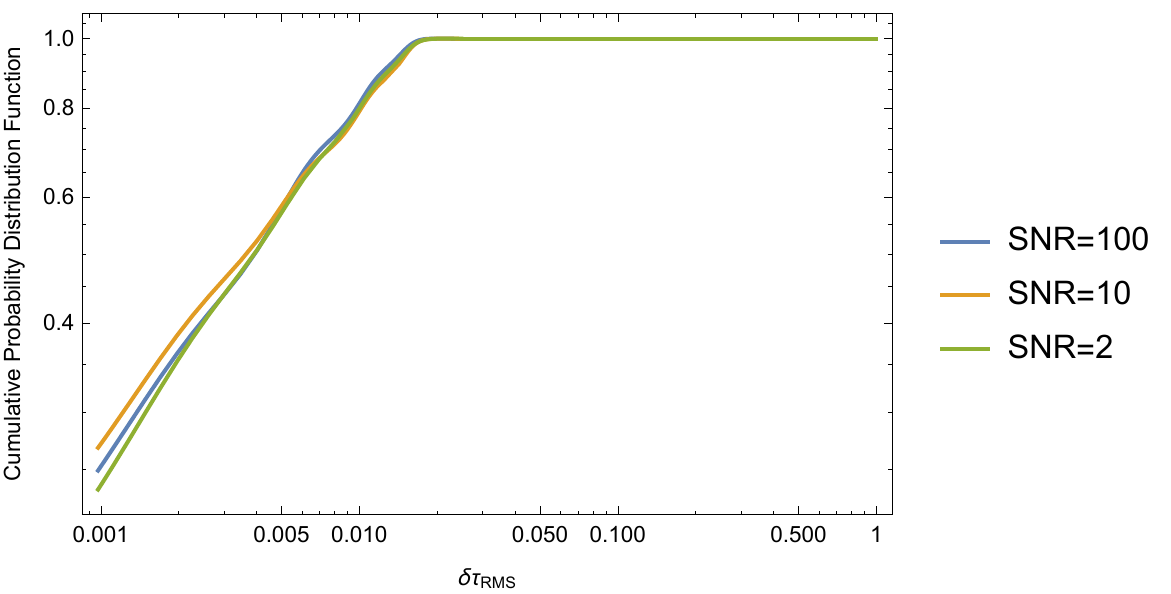}
\caption{This figure shows very minimal changes in the RMS time delay error across a very wide range of signal-to-noise ratios (see eq. \eqref{eq:ellipticdetectorresponse}) and fractional un-modeled signals (capped here at $\frac{1}{SNR}$; see eq. \eqref{eq:ellipticums}). Future work will determine whether these minimal changes persist when the parameter space is sampled more densely.}
\label{fig:ellipticsnrtauplot}
\end{figure*}

\begin{figure*}
\centering
\includegraphics[scale=1.25]{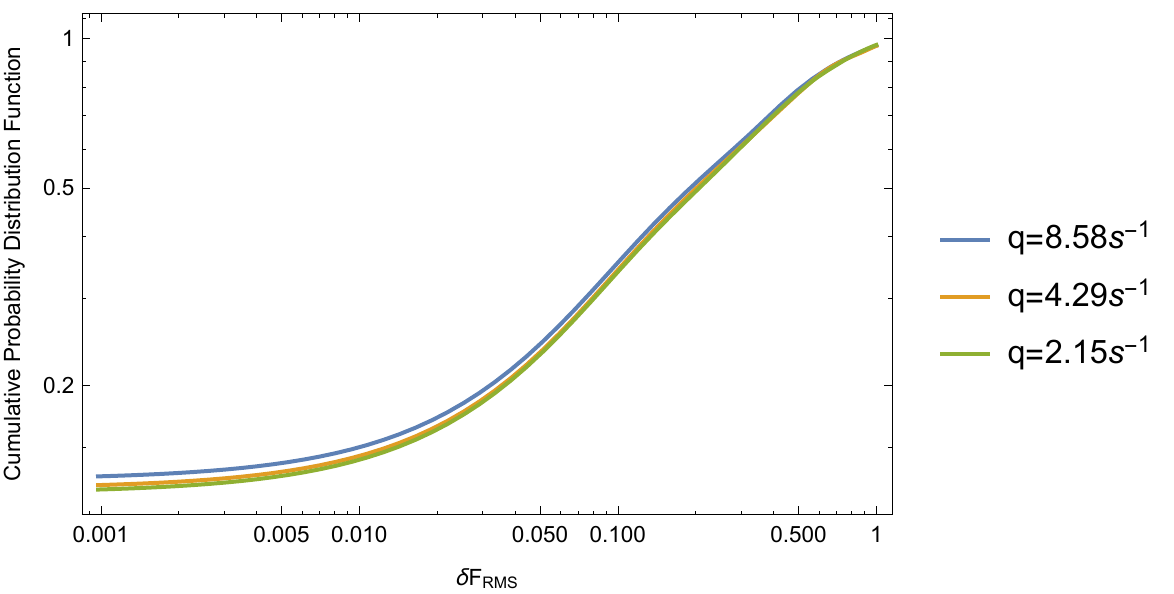}
\caption{This figure shows that the algorithm handles signals with different $q$-values equally well. These $q$-values correspond to signal lifetimes -- characterized by the time at which the signal amplitude drops to one-half its maximum value -- of $\frac{1}{4}$s, $\frac{1}{2}$s, and $1$s.}
\label{fig:ellipticlifetimeplot}
\end{figure*}

\begin{figure*}
\centering
\includegraphics[scale=1.25]{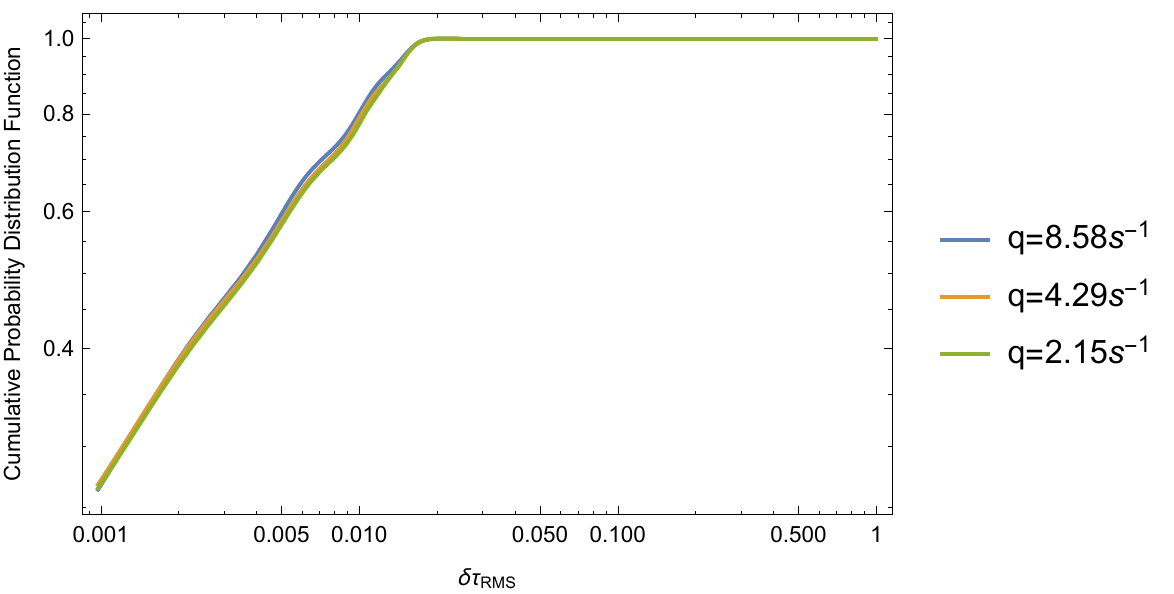}
\caption{This figure shows that the algorithm handles signals with different $q$-values equally well. These $q$-values correspond to signal lifetimes -- characterized by the time at which the signal amplitude drops to one-half its maximum value -- of $\frac{1}{4}$s, $\frac{1}{2}$s, and $1$s.}
\label{fig:ellipticlifetimetauplot}
\end{figure*}

To give a sense of the method's capability to give good results over a range of parameter values within the larger parameter space, figs. \ref{fig:ellipticfreqplot}, \ref{fig:ellipticsnrplot}, and \ref{fig:ellipticlifetimeplot} show the cumulative probability distribution function of $\delta F_{rms}$ over the frequencies, SNR/un-modeled signal fractions, and lifetimes indicated, while figs. \ref{fig:ellipticfreqtauplot}, \ref{fig:ellipticsnrtauplot}, and \ref{fig:ellipticlifetimetauplot} show the cumulative probability distribution function of $\delta \tau_{rms}$ over the same parameter sets. An ideal algorithm would give equivalent results (i.e., the same CPDF) across all possible frequencies, SNRs, and signal lifetimes, indicating that it could perform equally well throughout the entire parameter space. The algorithm's performance in determining $\tau$ is relatively close to this ideal. However, as seen in \cite{mcclain2018using}, higher frequency signals and higher maximum levels of noise and un-modeled signal reduce the accuracy of the method in finding beam $F_+$ and $F_\times$ at the two sites. Though the algorithm suffers at higher frequencies -- an effect likely due to the fact that even small non-zero values $\delta \tau$ are amplified by the frequency when determining the mismatch of eq. \eqref{eq:ellipticdetectorresponse} -- fig. \ref{fig:ellipticfreqplot} shows that the method retains predictive power throughout the frequency range of these simulations. Fig. \ref{fig:ellipticsnrplot} shows that the algorithm proves surprisingly resilient to high levels of noise and un-modeled signal. Though similar results were obtained in \cite{mcclain2018using} for circularly polarized GWs, this unusually high tolerance is -- as discussed above -- likely due to the relatively coarse sampling of the parameter space that has been used in these simulations; I do not expect this level of resilience when the parameter space is sampled more densely. Only signal lifetimes show the ideal behavior: fig. \ref{fig:ellipticlifetimeplot} indicates that the impact of q-value/signal lifetime on the method's accuracy is negligible.

\begin{figure*}
\centering
\includegraphics[scale=1.25]{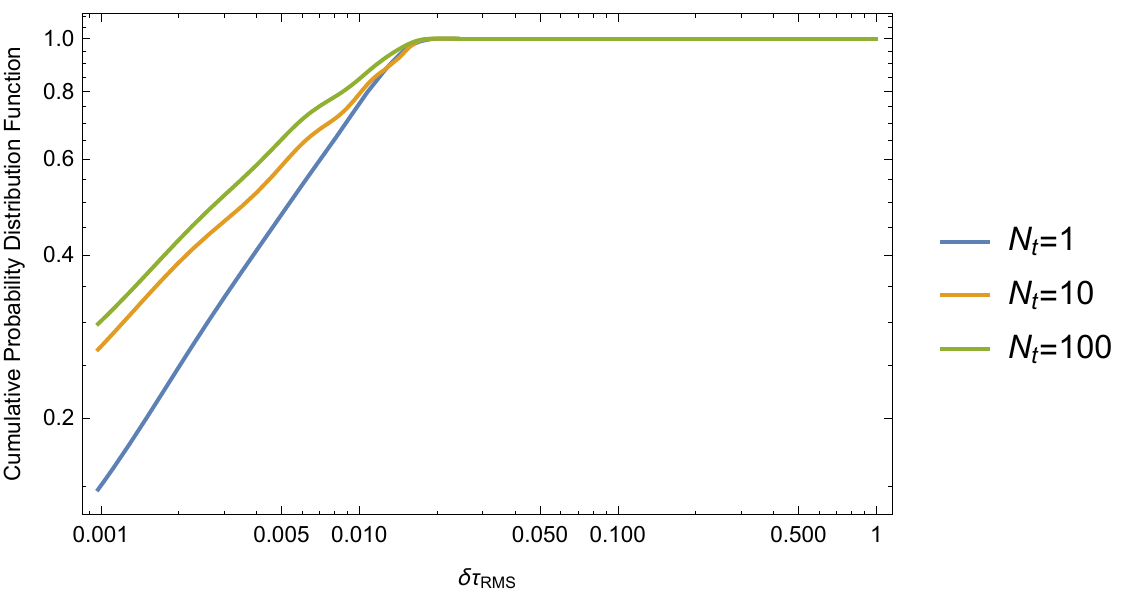}
\caption{This figure shows the substantial differences in fit RMS time delay error when the time portion of the parameter space is sampled more densely. See the caption of fig. \ref{fig:elliptictsampleplot} for more details.}
\label{fig:elliptictsampletauplot}
\end{figure*}

\begin{figure*}
\centering
\includegraphics[scale=1.25]{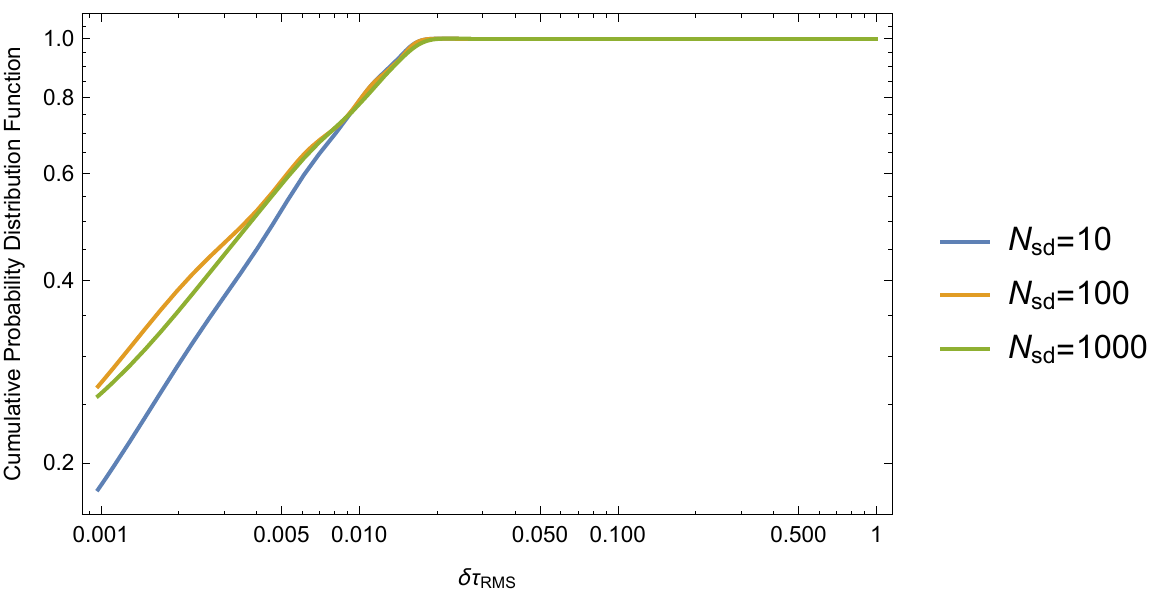}
\caption{This figure shows the notable differences in RMS time delay error across the different numbers of source direction samples. As with the beam pattern data, this indicates that the source directions are being appropriately sampled relative to times and GW amplitude combinations in the reference scenario. See section \ref{sec:results} for more details.}
\label{fig:ellipticanglestauplot}
\end{figure*}

\begin{figure*}
\centering
\includegraphics[scale=1.25]{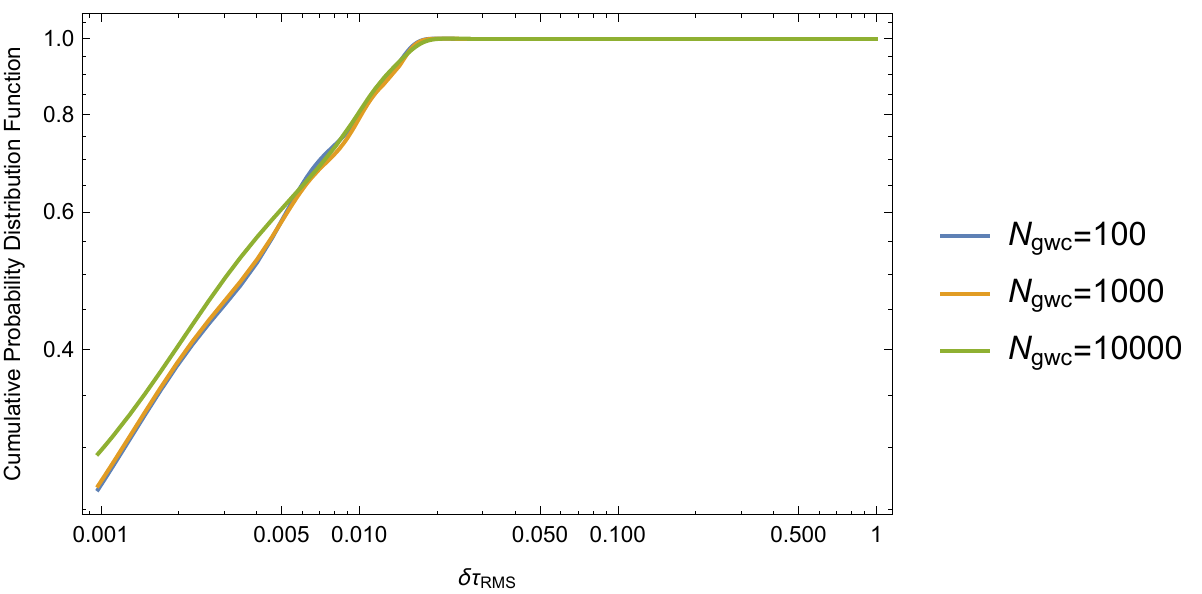}
\caption{This figure shows the very small difference in RMS time delay error across the different numbers of GW amplitude combinations. Like the beam pattern response data, this indicates both that GW amplitude combinations are being over-sampled relative to times and source directions in the reference scenario, but also that the sampling density would need to be significantly higher to produce high-accuracy fit results. See section \ref{sec:results} for more details.}
\label{fig:ellipticgwctauplot}
\end{figure*}

To give a sense of how the algorithm might fare when using higher sampling densities within this parameter space, I have done simulations that increase or decrease one of the three sampling densities while leaving the other two at their baseline levels; see figs. \ref{fig:elliptictsampleplot}, \ref{fig:elliptictsampletauplot}, \ref{fig:ellipticanglesplot}, \ref{fig:ellipticanglestauplot}, \ref{fig:ellipticgwcplot}, and \ref{fig:ellipticgwctauplot}. Here the ideal behavior is quite different from the three cases examined above: an ideal algorithm would show substantial improvement in the CPDF (an upward shift of the left-most portions of the graph) when any of its sampling densities is increased, indicating that the algorithm produces more accurate results at higher sampling densities. Similarly, the ideal algorithm would show a substantial worsening of the CPDF (a downward shift of the left-most portions of the graph) when any of its sampling densities is decreased, indicating that the ratio of sampling densities used in the algorithm is appropriate; i.e., there are no regions of the parameter space that are being oversampled relative to the others. Finally, an ideal algorithm would retain predictive power -- and lower values of $\delta F_{rms}$ and $\delta \tau_{rms}$ -- even at lower sampling densities. My results show that the algorithm does retain predictive power -- and continues to show smaller values of the key performance indicators $\delta F_{rms}$ and $\delta \tau_{rms}$ as compared to the single-best-fit or random fit methods -- even at very low sampling densities. Also, the method's average $\delta F_{rms}$ and $\delta \tau_{rms}$ values improve by significant amounts as any one of the sampling densities is increased. Increasing the time ($N_t$) and source direction ($N_{sd}$) sampling densities show a greater improvement in these key performance indicators than increasing the GW amplitude combination sampling density $N_{gwc}$. Similarly, decreasing the GW amplitude combination sampling density decreases the key performance indicators substantially less than decreasing either the time or the source direction sampling densities. Taken together, these results suggest that the ideal ratio of sampling densities is different from the baseline ratio used in these simulations \emph{at the particular level of accuracy achieved in this analysis}. The fact that the far-left portion of the CPDF does rise considerably at very high gravitational wave amplitude combination sampling densities shows that this sampling density may actually be too low if our target accuracy is quite high. Determining the ideal ratio of sampling densities in these kinds of different scenarios will be an important consideration in future work.

As was concluded in \cite{mcclain2018using}, more extensive computational testing with a set of networked cores is still necessary to prove that the algorithm will continue to perform well once massively parallelized. The current simulation results continue to show promise, suggesting that the time and computing resources necessary for this more extensive testing would be well spent. 

\section{Discussion and conclusions}
\label{sec:discussion}

The percent increase in accuracy of the best-performing weighting function of eq. \eqref{eq:ellipticweightedQ} over that of eq. \eqref{eq:ellipticminQ} ranges from $14 \%$ to $26 \%$ in the scenarios tested, with the computational load increasing by only as much as $2.4 \%$ (and in some cases much less) due to the time it takes the algorithm to perform the additional weighting required by eq. \eqref{eq:ellipticweightedQ} that is not required to compute eq. \eqref{eq:ellipticminQ}. Happily, both the higher fractional improvements in accuracy and the lower fractional increases in computational load occur at higher sampling densities -- i.e., in the most realistic scenarios tested. For reference, the baseline parameter values  \(q = 4.29 \, \text{s}^{-1} \), \( f =100 \, \text{Hz} \), \( \text{SNR} \geq 10 \), the parameters \( u_1 \) through \( u_5 \) of eq. \eqref{eq:ellipticums} all capped at \( u_{\text{max}} = 1/10 \), $N_t = 10$ random times, $N_{sd} = 100$ random source directions, and $N_{gwc} = 1000$ random gravitational wave amplitude combinations yield a median accuracy increase of $17 \%$ for a median computational load increase of $0.1 \%$. 

There are two main results that suggest that the method remains promising in the larger parameter space used in this analysis. First, the weighting algorithm seems to perform almost equally well under small changes of the form of the weighting function, in contrast to the results of \cite{mcclain2018using}. Second, the weighting algorithm continues to show smaller average values of $\delta F_{rms}$ and $\delta \tau_{rms}$ than the single- and random- best-fit methods across a wide range of frequencies, q-values/signal lifetimes, and SNRs. In fact, by every calculated metric the revised weighting algorithm performs better in the larger parameter space than in the reduced parameter space of \cite{mcclain2018using}.

To re-iterate the results of section \ref{sec:results}, the simulations run for this analysis indicate that the sharply-decreasing exponential weighting of eq. \eqref{eq:ellipticweightedQ} with $n = 2$ (i.e., a Gaussian weighting function) outperforms the single best fit weighting of eq. \eqref{eq:ellipticminQ} in every scenario, including the presence of substantial noise and/or un-modeled signal. Though still not conclusive, the current analysis indicates that use of a revised weighting algorithm of the form of eq. \eqref{eq:ellipticweightedQ} is likely to produce substantial improvement over the algorithm of eq. \eqref{eq:ellipticminQ} in a realistic GW source localization scenario. Indeed, in the larger parameter space it is not even necessary to quantify the amount of noise in advance of applying the weighting algorithm to achieve near-optimal results (as was the case in \cite{mcclain2018using}).

To achieve this improved output in the context of an algorithm like LALInference, it would be necessary to use a finer mesh of parameter values, but the computational cost of doing so is vastly greater than that of implementing the revised weighting algorithm. Indeed, in larger parameter spaces the savings in computational load is much greater than in the situation referenced in \cite{mcclain2018using}. To establish a baseline, first consider the cost of obtaining a particular result with the weighting function of eq. \eqref{eq:ellipticminQ}. Using this weighting function does not require us to calculate the large array of weighted \( Q \) values that we must have if we wish to use the weighting function of eq. \eqref{eq:ellipticweightedQ}; the computational cost of calculating this large array represents the additional computational load of the revised weighting algorithm. In the case of the monochromatic sine-Gaussian model (much simplified compared to LALInference, with only eight parameters), using the $n=2$ weighting algorithm on a signal of frequency $f = 100 \text{Hz}$ with noise and un-modeled signal each less than $1/10$ of the signal value is expected to produce about a \( 17 \% \) median improvement in accuracy based on the results of current simulations, while increasing the computational load of the model by less than \( 1 \% \) based on direct measurement of simulation runtimes. On the other hand, to achieve this same average improvement in accuracy using a finer mesh of parameters values requires (in a naive, best-case scenario) using a \( 17 \% \) finer mesh, which in turn increases the computational load of the model by \( \approx 250 \% \): every large array in the simulation grows by that same factor, and the computational cost of the simulation is dominated by a few computations involving those large arrays. In actual simulations, achieving a $17 \%$ increase in accuracy increases the computational load of the algorithm by $ 340 \%$ to $1300 \% $, depending upon which parts of the parameter space are sampled more density to generate the improvement. As in \cite{mcclain2018using}, the recommended strategy would therefore be to use the finest parameter mesh possible with a given set of computational resources, and then to implement the revised weighting algorithm to achieve a final \( 17 \% \) improvement in accuracy without much attendant increase in computational load.

The continued -- indeed increasing -- success of the new weighting algorithm in the more realistic scenarios tested in this analysis suggests that it may be worthwhile to invest the time and resources necessary to test this algorithm in real-world gravitational wave detection scenarios. To do this, two additional steps are necessary. First, the algorithm itself must be re-written in a more computationally efficient language to run in parallel on an arbitrary number of cores. Second, a very large amount of compute-time must be spent running the revised code on a highly parallelized network of CPUs to check the performance of the algorithm in many simulations of realistic GW search scenarios. It is my hope that this follow-up analysis will provide the necessary groundwork for these next steps.

\bibliographystyle{unsrt}
\bibliography{gw.bib}

\end{document}